\documentclass[aps,prl,reprint,amsmath,amssymb,twocolumn,superscriptaddress,notitlepage,groupedaddress]{revtex4-1}
\usepackage{graphicx}
\usepackage{dcolumn}
\usepackage[dvipsnames*,svgnames]{xcolor}
\usepackage{verbatim}
\usepackage{listings}
\usepackage{mathtools}
\usepackage{amsmath}
\usepackage{chngcntr}
\usepackage{bm}
\usepackage{url}

\usepackage[breaklinks]{hyperref}
\usepackage{breakurl}
\usepackage{tabularx}
\usepackage{soul}
\hypersetup{
colorlinks,
linkcolor={blue!100!black},
citecolor={blue!50!black},
urlcolor={blue!80!black}}
\begin{document}
\title{Thermal Anisotropy Enhanced by Phonon Size Effects in Nanoporous Materials}
\author{Giuseppe Romano}
\email{romanog@mit.edu}
\affiliation{Department of Mechanical Engineering, Massachusetts Institute of Technology, 77 Massachusetts Avenue, Cambridge, MA 02139, USA}
\author{Alexie M. Kolpak}
\affiliation{Department of Mechanical Engineering, Massachusetts Institute of Technology, 77 Massachusetts Avenue, Cambridge, MA 02139, USA}
\begin{abstract}
While thermal anisotropy is a desirable materials property for many applications, including transverse thermoelectrics and thermal management in electronic devices, it remains elusive in practical natural compounds. In this work, we show how nanoporous materials with anisotropic pore lattices can be used as a platform for inducing strong heat transport directionality in isotropic materials. Using density functional theory and the phonon Boltzmann transport equation, we calculate the phonon-size effects and thermal conductivity of nanoporous silicon with different anisotropic pore lattices. Our calculations predict a strong directionality in the thermal conductivity, dictated by the difference in the pore-pore distances, i.e. the phonon bottleneck, along the two Cartesian axes. Using Fourier's law, we also compute diffusive heat transport for the same geometries obtaining significantly smaller anisotropy, revealing the crucial role of phonon-size effects in tuning thermal transport directionality. Besides enhancing our understanding of nanoscale heat transport, our results demonstrate the promise of nanoporous materials for modulating anisotropy in thermal conductivity.\end{abstract}
\maketitle

\section{Introduction}
Designing materials to achieve desired thermal properties is pivotal for many modern applications, including thermoelectrics~\cite{CRC1995} and heat management in electronic devices~\cite{pop2010energy}. Use of nanostructures has led to the achievement of new extremes in both high and low thermal conductivities relative to natural materials ~\cite{kim2007nanostructuring}. The degree of transport suppression is dictated by the relationship between the characteristic length of the material and the dominant phonon mean free path (MFP)~\cite{chenbook}. For example, recent experiments have demonstrated very low thermal conductivities in thin films~\cite{Venkatasubramanian2001}, nanowires~\cite{boukai2008silicon,hochbaum2008enhanced}, and porous materials~\cite{song2004thermal,lee2015ballistic,Tang2010,Hopkins2011,kargar2015acoustic,de2011temperature}, illustrating the effectiveness of boundary engineering in tuning thermal transport~\cite{Romano2014}.

 While suppression of phonon transport in nanostructured materials has recently attracted a great deal of attention, little is known about the directionality of thermal conductivity in such materials. Tailoring thermal anisotropy, i.e., directional-dependent heat transport, can be useful for many applications, including transverse thermoelectrics, in which electron and phonon flows are orthogonal to each other~\cite{goldsmid2011application}. In-plane anisotropic thermal conductivities have been demonstrated in layered two-dimensional (2D) materials such as arsenene~\cite{zeraati2016highly} and phospherene~\cite{fei2014enhanced}, which also exhibit promising thermoelectric properties. In these systems, the in-plane directionality of heat transport arises from the puckered nature of the structures. In general, however, materials with both native anisotropic thermal transport and practical thermoelectric efficiencies remain elusive.  The wide range of pore sizes and configurations that have been demonstrated~\cite{Romano2014,romano2015,Yang2013,Romano2012} suggest that nanoporous materials are a good platform for artificially inducing and tuning thermal anisotropy in isotropic materials. In this work, we investigate this possibility using nanoporous silicon as a text case. Using our recently developed approach based on the Boltzmann Transport Equation (BTE)~\cite{romano2015}, we calculate the phonon-size effects and the thermal conductivity tensor in nanoporous silicon with anisotropic pore lattices. We consider three different pore arrangements: the configuration with fixed pore size, the case where the pore size is adjusted to keep the porosity fixed and the case where both pore size and porosity are kept fixed. For all the configurations, we observe a significant thermal conductivity anisotropy. These findings can be explained in terms of the directionality of the phonon bottleneck, represented by the pore-pore distance along the applied temperature gradient. On the other side, simple Fourier's law simulations for the same geometries predict weak anisotropy, revealing the importance of phonon-size effects in enhancing thermal transport directionality.

\begin{figure}[h!]
\begin{center}
\includegraphics[width=1\columnwidth ]{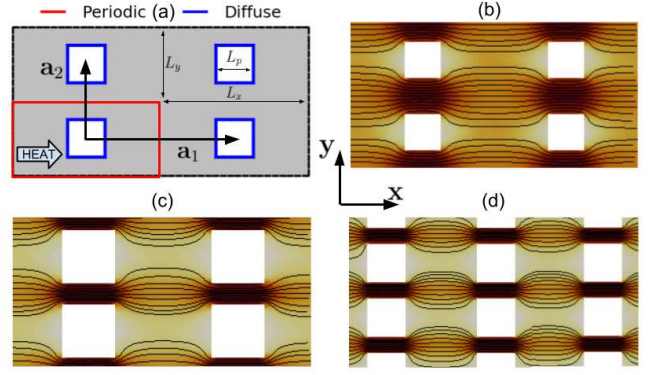}
\caption{(a) The simulation domain comprises one single square pore. Periodic boundary conditions are applied both along \textit{x}- and \textit{y}-directions. The walls of the pores are assumed to be diffusive. Magnitude of thermal flux when a temperature gradient is applied along the \textit{x}-direction for a shape ratio $r$ = 2 for the case with (a) fixed pore size (b) fixed porosity and (c) fixed pore size and porosity. For all the cases illustrated, $\phi=$ 0.25 and $L=$ 10 nm.}\label{Fig:10}
\end{center}
\end{figure}

Thermal transport simulations are performed over a unit-cell with size $L_x$x$L_y$ comprising a single square pore of size $L_p$, as shown in Fig.~\ref{Fig:10}-a. The pore lattice is identified by the orthogonal lattice vectors $\mathbf{a}_1 = L_y r \mathbf{\hat{x}}$ and $\mathbf{a}_2 =L_y \mathbf{\hat{y}}$, where $r$ is the shape factor, simply defined as $r=L_x/L_y$.  For $r$ = 1, the pore lattice is isotropic and the unit-cell is a square of side $L_x=L_y=L$ = 10 nm with pore size $L_p = L \sqrt{\phi}$ = 5 nm, where $\phi$ = 0.25 is the porosity, i.e. the ratio between the area of the pore and the area of the unit-cell. We will refer to the isotropic case as ``ISO''. For anisotropic pore lattices, we consider three cases of practical interests: fixed porosity (FP), fixed pore size (PS) and fixed porosity and pore size (FSP), as illustrated in Fig.~\ref{Fig:10}-b,-c, and -d, respectively. Anisotropic pore lattices, which correspond to $r>1$, are achieved by varying $L_x$ and adjusting $L_y$ and $L_p$ in order to achieve the conditions required for the specific configuration. The values for these parameters are described in table~\ref{Tab:10}. The thermal conductivity tensor is reconstructed by applying a difference of temperature $\Delta T_\alpha = $ 1 K along the Cartesian axis $\alpha$ and collecting heat flux along the same direction. Then, the application of Fourier's law gives $\kappa_{\alpha \alpha}=\frac{L_{\alpha}}{\Delta T}\int_{C_\alpha}\mathbf{J}(\mathbf{r},\Lambda,\Omega)\cdot\mathbf{n}dS$, where $C_\alpha$ is the boundary of the unit-cell with normal aligned with $\alpha$.

\begin{figure}[h!]
\begin{center}
\includegraphics[width=1\columnwidth ]{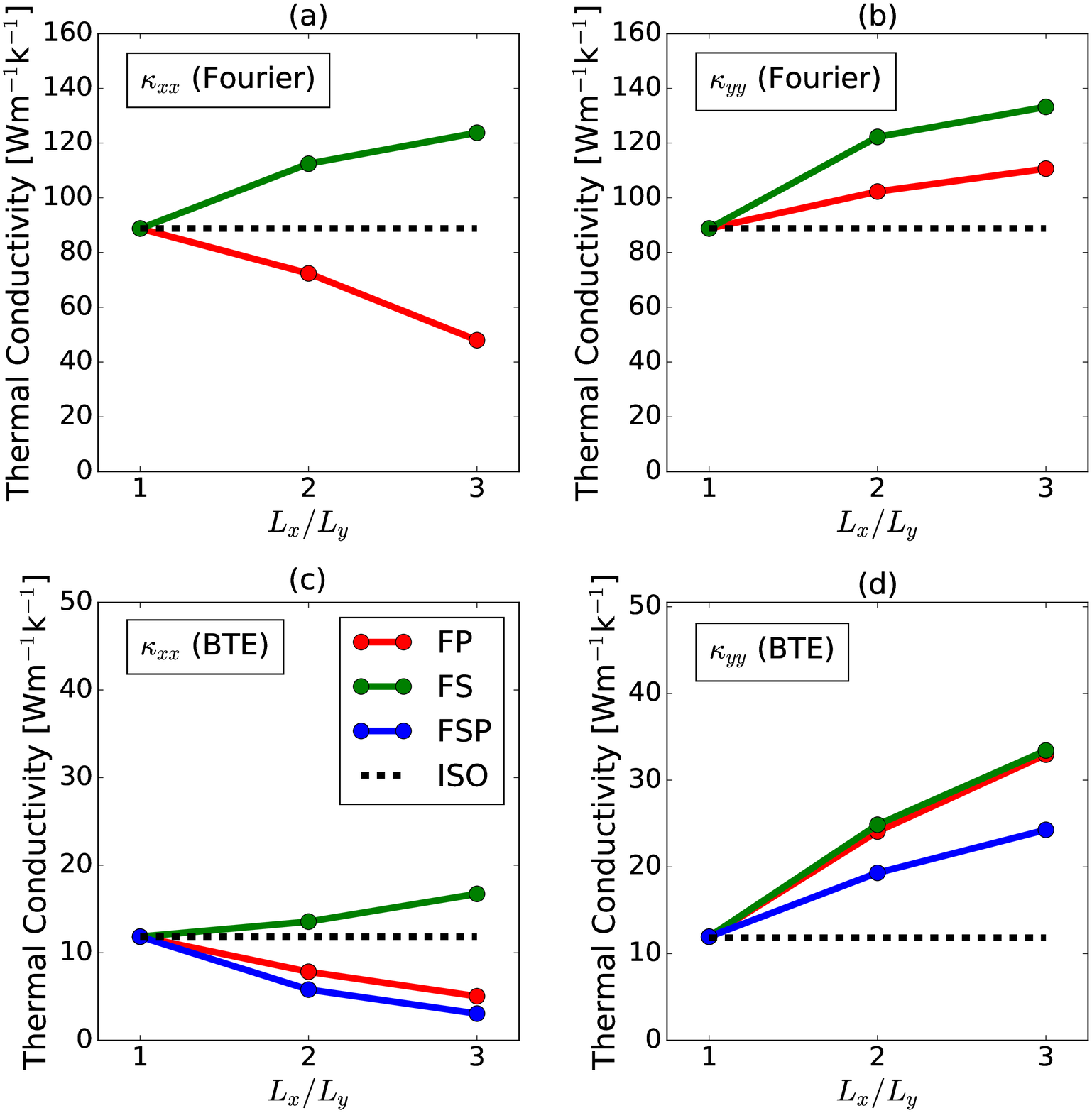}
\caption{Thermal conductivity tensor for different cases (FP, FS and FSP), shape ratios $r$ and with $L$ = 10 nm. The component (a) $\kappa^D_{xx}$ and (b) $\kappa^D_{yy}$ computed by Fourier's law. The quantities (c) $\bar{\kappa}_{xx}$ and (d) $\bar{\kappa}_{yy}$ computed by the MFP-BTE. In all the subplots, the black-dotted line refers to the ISO case.}\label{Fig:20}
\end{center}
\end{figure}

\begin{figure}[h!]
\begin{center}
\includegraphics[width=1\columnwidth ]{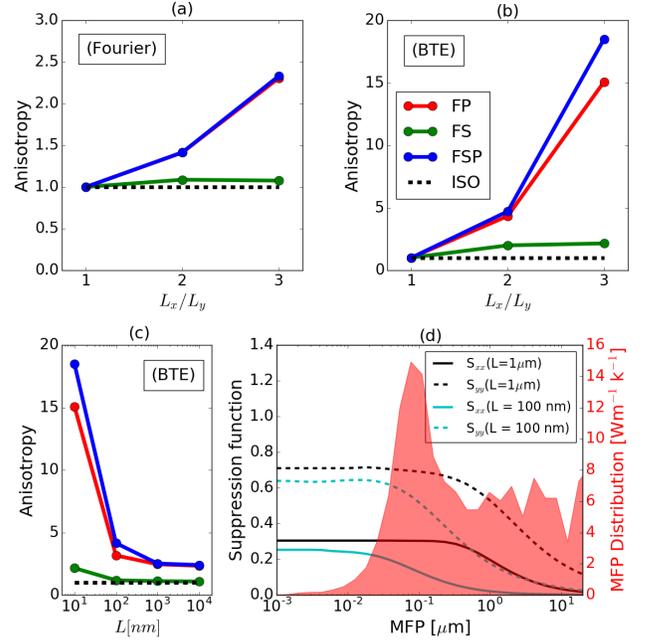}
\caption{The anisotropy (a) $A^D$ and (b) $A^B$ for different cases (FP, FS and FSP), shape ratios $r$ and with $L=$ 10 nm. (c) The anisotropy computed by BTE and for different scales $L$. (d) The phonon suppression functions $\bar{S}_{\alpha \alpha}(\Lambda)$ for $L=$ 100 nm and $L=$ 1 $\mu$m. In the background, the bulk MFP distribution used as input to the BTE.}\label{Fig:30}
\end{center}
\end{figure}

\begin{table}
\caption{Geometrical parameters for the different configurations in relation to Fig.~\ref{Fig:10}.}
\label{Tab:10}
\begin{center}
\begin{tabular}{ |c||c|c|c|}
\hline
 &Same size&Same porosity&Same size/porosity\\ \hline
\hline
$L_x$&$L\,r$&$L\,r$&$L\,\sqrt{r}$\\ \hline
$L_y$&$L$&$L$&$L/\sqrt{r}$\\ \hline
$L_p$&$L\sqrt{\phi}$&$L\sqrt{r\phi}$&$L\sqrt{\phi}$\\ \hline
\end{tabular}
\end{center}
\end{table}
We first investigate heat transport in absence of phonon size effects, by using a finite-volume solver for the standard diffusive equation $\kappa_{bulk}\nabla^2 T(\mathbf{r})=0$, where $\kappa_{bulk}$ is the bulk thermal conductivity ($\sim$155 Wm$^{-1}$K$^{-1}$ for Si~\cite{glassbrenner1964thermal}) and ``D'' stands for ``diffusive'. For the ISO case, the Maxwell-Garnett theory predicts the reduction in the diffusive heat transport $\kappa^D_{xx}=\kappa^D_{yy}\approx \kappa_{bulk}f(\phi)=\kappa_{bulk}(1-\phi)/(1+\phi)\approx$ 90 Wm$^{-1}$k$^{-1}$~\cite{nan1997effective}. In regard to the anisotropic cases, the FP and FSP cases lead to the same thermal conductivities because the two configurations differ only by a scale factor, a parameter to which diffusive heat conduction is insensitive. In this case, for $r>1$, the pore-pore distance along $\mathbf{\hat{x}}$ shrinks leading to a decrease in $\kappa^D_{xx}$, while $\kappa^D_{yy}$ increases because of the widening of the distance between the pores along $\mathbf{\hat{y}}$. These opposite trends lead to an anisotropy, defined by $A^D=\kappa^D_{yy}/\kappa^D_{xx}$, of $\sim$ 2.4 for $r$ = 3. On the other side, for the FS case, both $\kappa^D_{xx}$ and $\kappa^D_{yy}$ increase with $r$ because of the decrease in the porosity. In this case, the anisotropy remains relatively small.  The values of $\kappa^D_{xx}$ and $\kappa^D_{yy}$ for different $r$ are shown in Fig.~\ref{Fig:20}-a and -b, respectively, while the values of $A^D$ are plotted in Fig.~\ref{Fig:30}-a.

Let us analyze the effects of phonon-boundary scattering on the anisotropy. Phonon size effects are computed by solving the Mean-Free-Path BTE (MFP-BTE), which reads as~\cite{romano2015} \begin{equation}
\begin{split}\label{Eq:1}
\Lambda \mathbf{s}(\Omega)\cdot\nabla T(\mathbf{r},\Lambda,\Omega) +T(\mathbf{r},\Lambda,\Omega)=\\=\gamma \int_0^\infty \frac{K(\Lambda)}{\Lambda^2}<T(\mathbf{r},\Lambda,\Omega)>d\Lambda,
\end{split}
\end{equation}
where $T(\mathbf{r},\Lambda,\Omega)$ is the effective temperature associated with phonons with MFP $\Lambda$ and direction $\mathbf{s}(\Omega)$, $<.>$ is an angular average and $\gamma$ is a material property, defined as $\gamma = \left[\int_0^\infty K(\Lambda)/\Lambda^2 d\Lambda\right]^{-1}$. The term $K(\Lambda)$, namely the only input to our model other than the material's geometry, is the bulk MFP distribution, computed by the density functional theory~\cite{broido2007intrinsic,esfarjani2011heat}. Eq.~\ref{Eq:1} is discretized in space by means of the finite-volume method~\cite{romano2011multiscale} and in angular domain by the discrete ordinate method~\cite{abe1997derivation}. Periodic boundary conditions are applied along both $\mathbf{\hat{x}}$ and $\mathbf{\hat{y}}$, while the walls of the pores are assumed to scatter phonon diffusively. This condition is achieved by enforcing zero normal flux along the pore's wall, which leads to the following condition on the boundary temperature~\cite{romano2016temperature} \begin{equation}
\begin{split}\label{Eq:2}
T_b=-\frac{\int_0^\infty \int_{\mathbf{s}(\Omega)\cdot\mathbf{n} \geq 0 } \mathbf{J}(\mathbf{r},\Lambda,\Omega) \cdot \mathbf{n}\, d\Omega d\Lambda}{ \int_{\mathbf{s}(\Omega)\cdot\mathbf{n}<0}\frac{K(\Lambda)}{\Lambda^2}\mathbf{s}(\Omega)\cdot\mathbf{n}\,d\Omega d\Lambda},
\end{split}
\end{equation}where $\mathbf{n}$ is the normal to the pore's surface and $\mathbf{J}(\mathbf{r},\Lambda,\Omega)=\frac{K(\Lambda)}{\Lambda}T(\mathbf{r},\Lambda,\Omega)\mathbf{s}(\Omega)$ is the thermal flux. 
The presence of phonon-boundary scattering results in a strong reduction in thermal transport. For example, for the ISO case, $\kappa^B_{xx} = \kappa^B_{yy} = 6$ Wm$^{-1}$K$^{-1}$ (``B'' stands for ``BTE''), less than one order of magnitude smaller than that of the bulk counterpart~\cite{romano2016temperature}. In the subsequent analysis, in order to focus on phonon-size effects we compute the quantities $\bar{\kappa}_{\alpha \alpha}=\kappa_{bulk}\kappa_{\alpha \alpha}^B /\kappa^D_{\alpha\alpha}$. As shown in Fig.~\ref{Fig:20}(c), for the FS case $r>1$ leads to a modest increment in $\bar{\kappa}_{xx}$ and a significant increase in $\bar{\kappa}_{yy}$, leading to the anisotropy $A^B\approx$ 2.8 for $r$ = 3. This high anisotropy compared to that of the diffusive counterpart, can be explained in terms of the characteristic length, $L_c$. When heat transport is dominated by boundaries, the thermal conductivity goes with $L_c$~\cite{chenbook}, i.e. $\bar{\kappa}_{xx}\approx Kn^{-1}$, where $Kn$ is the Knudsen number, defined as $\Lambda/L_c$ and $\Lambda$ is the dominant MFP~\cite{chenbook}. For anisotropic pore lattices, the characteristic length $L_c$ depends on the direction of the applied temperature. For $r >$  1, phonons travelling along $\mathbf{\hat{y}}$ experience a larger $L_c$ than that related to phonons traveling along $\mathbf{\hat{x}}$. Hence, directionality in $L_c$ translates in enhanced anisotropy in thermal conductivity. According to this argument, for the FP and FSP cases, we obtain a decrease in $\bar{\kappa}_{xx}$ and an increase in $\bar{\kappa}_{yy}$ that reflects the trend of $L_c$ along $\hat{\mathbf{x}}$ and $\hat{\mathbf{y}}$, respectively. The highest anisotropy, $A^B\sim 18$, is obtained for the FSP case. From Fig.~\ref{Fig:10}, we note that phonons contributing to $\bar{\kappa}_{xx}$ travel mainly through direct paths, as a consequence of the small bottleneck. This ``phonon focusing'' effect can be suitable for coupling a heat source with a nanowire, as proposed in a recent study~\cite{anufriev2016heat}.

In order to provide predictions for material systems that are currently within experimental reach, we analyze the scale dependence of the anisotropy for all the three cases. While for the FS case the anisotropy becomes negligible for $L >$ 10 nm, the FP and FSP cases exhibit $A^B\sim 5$ for a scale as large as $L$ = 100 nm and approach the diffusive limit for larger scales, as shown in Fig.~\ref{Fig:30}. This result can be better analyzed in terms of the phonon suppression function, a quantity that describes phonon suppression for each MFP, defined as~\cite{romano2015} \begin{equation}
\begin{split}\label{Eq:3}
S_{\alpha \alpha}(\Lambda) = \frac{L_\alpha}{A_{\mathrm{hot}}\Delta T_{\alpha}\Lambda}\int_{\mathrm{hot}}<T(\mathbf{r},\Lambda,\Omega)\mathbf{s}(\Omega)\cdot \mathbf{n}>dA,
\end{split}
\end{equation}where $A_{\mathrm{hot}}$ is the area of the hot contact and $\mathbf{n}$ is its normal. The term $S(\Lambda)_{\alpha \alpha}$ provides the calculation of the thermal conductivity via $\kappa_{\alpha \alpha} = \int_0^{\infty}K(\Lambda) S_{\alpha \alpha}(\Lambda) d\Lambda$ and maps the bulk MFP distribution into that of the nanostructure. We narrow our analysis to the FSP case, which is the case of most interest. In fact, in a practical implementation, the reachable pore size is determined by the experiment limitations. On the other side, keeping the porosity fixed guarantees higher control in thermoelectric applications, where electrons, which mainly travel diffusively, are affected largely by the porosity rather than the actual pore configurations~\cite{romano2016tuning}. In order to focus on size effects, again, we compute the quantity $\bar{S}_{\alpha \alpha}(\Lambda) =\left(\kappa_{bulk}/\kappa^D_{\alpha \alpha}\right)S_{\alpha \alpha}(\Lambda)$. In Fig.~\ref{Fig:30}-d, we plot $\bar{S}_{xx}(\Lambda)$ and $\bar{S}_{yy}(\Lambda)$ for $L$ = 100 nm and $L = 1 \mu$m. In the background, we plot the bulk MFP distribution. We note that for the largest scale, $\bar{S}_{xx}(\Lambda) \approx \bar{S}_{yy}(\Lambda)$ for most of the spectrum while they differ significantly only for $\Lambda>2-3\mu$m, where the contribution to thermal conductivity is relatively small. On the other side, for $L = $ 100 nm, $\bar{S}_{xx}(\Lambda)$ is always higher than $\bar{S}_{yy}(\Lambda)$ because of the small $L_c$ with respect to the bulk dominant MFPs, giving rise to a significant anisotropy.

In summary, we have computed the thermal conductivity tensor of porous Si with anisotropic pore lattices, demonstrating that phonon size effects can induce significant anisotropy. The calculations were based on a recently developed method, which computes phonon-size effects with no input parameters other than the material's geometry and crystal structure. As these results apply when the constituting material is either isotropic or anisotropic, our findings suggest a practical route to artificially induce or suppress thermal transport directionality, an appealing capability for modern applications such as thermoelectrics and heat management in electronic devices. 
\section{Acknowledgments}
The authors thank Dr. Alexei Maznev for helpful discussions. Research supported as part of the Solid-State Solar-Thermal Energy Conversion Center (S3TEC), an Energy Frontier Research Center funded by the U.S. Department of Energy (DOE), Office of Science, Basic Energy Sciences (BES), under Award DESC0001. 
\end{document}